\documentclass[10pt,letterpaper]{article}
\usepackage[top=0.85in,left=2.75in,footskip=0.75in]{geometry}

\pdfoutput=1

\usepackage{booktabs}
\usepackage[normalem]{ulem}
\usepackage{multirow}
\useunder{\uline}{\ul}{}

\usepackage{mathtools}

\usepackage[utf8x]{inputenc}

\usepackage{csquotes}

\usepackage{amsmath,amssymb}

\usepackage{changepage}

\usepackage[utf8x]{inputenc}

\usepackage{textcomp,marvosym}

\usepackage{cite}

\usepackage{nameref,hyperref}

\usepackage[right]{lineno}

\usepackage{microtype}
\DisableLigatures[f]{encoding = *, family = * }

\usepackage[table]{xcolor}

\usepackage{array}

\newcolumntype{+}{!{\vrule width 2pt}}

\newlength\savedwidth

\usepackage{setspace} 

\raggedright
\usepackage[aboveskip=1pt,labelfont=bf,labelsep=period,justification=raggedright,singlelinecheck=off]{caption}

\makeatletter
\renewcommand{\@biblabel}[1]{\quad#1.}
\makeatother

\usepackage{lastpage,fancyhdr,graphicx}
\usepackage{epstopdf}
\fancyheadoffset[L]{2.25in}
\fancyfootoffset[L]{2.25in}
\begin{document}
\vspace*{0.2in}

\begin{flushleft}
{\Large
\textbf\newline{Phase transition modelling of relapse in major depressive disorder: Developing and reflecting on an interdisciplinary conceptual translation} 
}
\newline
\\
Marieke M. Glazenburg\textsuperscript{1*},
Luca Consoli\textsuperscript{2},
Alix McCollam\textsuperscript{3$\dagger$}
\\
\bigskip
\textbf{1} Department of Bionanoscience, Delft University of Technology, 2629 HZ Delft, The Netherlands
\\
\textbf{2} Institute for Science in Society, Radboud University Nijmegen, 6500 GL, The Netherlands
\\
\textbf{3} High Field Magnet Laboratory (HFML-EMFL) \& Radboud Interfaculty Complexity Hub, Radboud University, 6525 ED Nijmegen, The Netherlands
\\
\bigskip

%
%





* m.m.glazenburg@tudelft.nl, $^\dagger$ a.mccollam@ru.nl

\end{flushleft}
\section*{Abstract}
A tipping point can be defined as an abrupt shift in the properties or behaviour of a system. Tipping points in complex systems from a wide variety of scientific disciplines have been compared to phase transitions in physics, but consistent methodology for modelling tipping points as phase transitions has been lacking. Here, we propose a systematic approach aimed at order parameter identification in systems outside of physics undergoing an apparent regime shift. Based on classical Landau theory, we assess the relatedness of a system's properties to the order parameter by means of two quantitatively operationalized criteria: (1) the presence of a significant level shift over the course of the tipping point and (2) increased fluctuations before the tipping point. We first demonstrate the feasibility of our method by applying it to a case study of a tipping point in major depressive disorder, resulting in a list of symptoms that are most likely to be closely related to the order parameter in this particular system. Subsequently, we probe the usefulness of our approach in the interdisciplinary context of complexity science by means of exploratory interviews with active scientists. Our results suggest a growing need for interdisciplinary methodologies in complex systems studies, to which the phase transition modelling we present could provide a valuable addition.


\section*{Introduction}
Over the past years, complexity science has been increasingly recognised for its potential to address intricate, societally relevant problems by making use of complex systems techniques from different scientific disciplines \cite{Manson_2001}. Complex systems can be defined as systems consisting of large numbers of mutually interacting components, causing collective behaviour that is not observed in the individual constituents separately \cite{Bar_2019}. Such systems are ubiquitous throughout science and appear in the context of for instance physics, biology, psychology and many other fields \cite{Whitesides_1999,Sole_2000,Olthof_2020}. Despite the differences in their nature and the scientific disciplines in which they are studied, many complex systems possess similar characteristics \cite{Boehnert_2018}, such as self-organizing properties \cite{Couzin_2003,Packard_1988}, dynamical or adaptive behaviour \cite{Holland_1992,Strogatz_1994} and emergent pattern formation \cite{Ball_2004,Nicolis_2012}. These parallels make it possible, in principle, to describe a wide variety of complex systems using similar models \cite{Bertalanffy_1968,Woldenberg_1979,Changizi_2001}. The translation of concepts and methods between disciplines, however, has practical and cultural barriers that need to be explored in the process of building a truly interdisciplinary field such as complexity science. Mixed method approaches to transdisciplinary models of complex system properties, integrating quantitative and qualitative assessment, can provide insight into the context, usefulness and potential acceptance of a model, as well as its feasibility \cite{Lorenc_2016,Papoutsi_2021,Fetters_2013,Cresswell_2011,Curry_2013}. This holistic evaluation of a model is especially relevant in societal contexts, where it can also inform the design of potential interventions to achieve a desired outcome\cite{Campbell_2007,Clark_2013}.

A phenomenon frequently observed in complex systems is the occurrence of abrupt, dramatic shifts in the properties or behaviour of a system, also known as tipping points \cite{Scheffer_2009}. Tipping points can occur as a consequence of only incremental variation of external parameters or even without discernible causal factors, often resulting in irreversible and catastrophic changes to the system's state. Examples vary from sudden population decline and epileptic seizures to stock market crashes and global climate change \cite{Dai_2012,Litt_2001,NRC_2007,Lenton_2008}. Due to the often profound consequences of tipping points for a system and its environment, their study has been particularly focused on obtaining predictive measures. Over the past years, certain classes of generic early warning signals have been discovered that precede some regime shifts in a wide variety of complex systems \cite{Scheffer_2012,Ives_1995,Dakos_2008,Carpenter_2006}.

From the perspective of condensed matter physics, a tipping point bears resemblance to a continuous phase transition. A phase transition occurs when a material moves from one (structured, magnetic, nematic, etc.) state to another under the variation of some external parameter (temperature, pressure, chemical composition, etc.) \cite{Hohenberg_2015}. The original phenomenological theory of phase transitions was formulated by Landau \cite{Landau_1937} and has since provided a point of departure for more sophisticated theories of increasingly exotic types of phase transitions. Landau theory involves the principle of universality, which refers to the assumption that a system's behaviour over a phase transition does not depend on its microscopic properties, but can instead be understood by a generic macroscopic description that does not rely on detailed system-specific knowledge. This implies that phase transitions with very different underlying mechanisms can still be described by the same (universal) framework. A central concept in this abstract description is the notion of an order parameter, a quantity capturing the evolution of the order of a system over the course of a phase transition \cite{Hohenberg_2015}. In general, an order parameter will change from zero in the disordered phase to some non-zero value in the ordered phase.

The apparent parallels between phase transitions in physics and regime shifts in other complex systems raise the question whether phase transition theory can be used to better understand tipping points outside of physics. Given the central position of the order parameter within the study of phase transitions in physics, identifying an appropriate order parameter for a specific tipping point provides a sensible starting point for this question. Landau's theory is formulated for equilibrium phase transitions, but the concept of an order parameter remains valid for non-equilibrium transitions \cite{Haken_1975}. It is therefore fully relevant for tipping points in dynamical systems, such as often found in health, climate and many of the most societally challenging problems.

The notion of an order parameter or a similar collective variable has already been referenced fairly frequently to help characterise regime shifts in complex systems \cite{Sole_1996,Schoner_1988,Davies_2011,Haken_1977,Thelen_1991}, but more general strategies have not yet been developed. Although this concept is pivotal in a study of phase transitions in generic complex systems, the methods used to identify an order parameter in a specific system are often unclear and seem to lack a systematic approach.

In order to make concrete advancements in this area, a case study was used. The selected case study originates from the context of behavioural science and consists of an extensive collection of ecological momentary assessment (EMA) data, first appearing in \cite{Wichers_2016}, concerning a patient recovering from major depressive disorder (MDD) who gradually discontinued their antidepressant medication. A sudden significant increase in depressive symptom severity was observed partway through the measurement period, which the authors classified as a tipping point. Complexity science and interdisciplinary methods are gaining traction in behavioural science and psychology \cite{Olthof_2020,Borsboom_2013}, which makes this a highly relevant context. Moreover, the level of detail of the data and the relatively clear indication of a tipping point made this dataset particularly well-suited for our purpose.

The aim of this study is therefore to take the first steps towards developing a systematic approach for identifying an order parameter in generic complex systems undergoing an apparent regime shift, based on quantitative analysis of the specific case study sketched above, where a tipping point has already been identified. As part of this aim, an exploratory qualitative examination of the potential scientific and/or practical value and corresponding limitations of using physics theory in an extra-disciplinary context will also be carried out.

This paper is organized as follows. First, we describe the dataset and the development of the order parameter identification methods, as well as the qualitative methods that were used. Then, results from both components of the study are presented. Subsequently, we develop a discussion of the results and their implications. We finish with final conclusions and an outlook.

\section*{Materials and methods}
\subsection*{Dataset}
The dataset used as a case study first appeared in \cite{Wichers_2016} and was later made publicly available \cite{Kossakowski_2017}. The data originate from a single participant, a 57 year old male, who had suffered from multiple depressive episodes and had been using antidepressant medication in the 8.5 years prior to the start of the experiment. During the experiment, his medication was gradually discontinued. Over the course of 238 days, the participant completed a total of 1476 questionnaires about his daily life experiences using ecological momentary assessment (EMA), which amounts to an average of 6.2 assessments per day. Only a total of 5 responses were aborted before completion.

Different items in the questionnaires were prompted at different time intervals: either multiple times a day, once a day or once a week. The weekly items concerned the 13 items from the Symptom Checklist Revised (SCL-90-R \cite{Derogatis_2010}) depression scale. The average score based on these 13 items represents the depressive symptom severity. The daily items are divided into six items collected each morning regarding quality of sleep and six items collected each evening regarding the quality of the day as a whole. The momentary items appeared in all 10 measurement prompts and were divided into the following categories: mood states, self esteem, social company, physical sensations, the activity at the moment of the assessment and the occurrence of events in between the assessments. 33 items were scored on a 7-point Likert scale, ranging from 1 (not) to 7 (very). 6 items (related to mood states and events) were measured on a different 7-point scale ranging from -3 (not) to 3 (very), because it was found that this range increased the variation that the participant reported in his responses. Items related to social company, activities and events that required a qualitative response were categorised numerically (e.g. ``I am with friends" was assigned value 30).

During the measurement period, the dosage of the participant's antidepressant was gradually reduced at a rate that was unknown to both the participant and the researchers. The experiment consisted of five phases: (1) a baseline measurement of 4 weeks, (2) a double-blind period without reduction of the antidepressant dosage of between zero and six weeks, (3) a double-blind period in which the dosage was gradually reduced to zero over the course of eight weeks, (4) a post-assessment period with constant zero dosage of eight weeks and (5) a follow-up period of twelve weeks. Around day 127 of the experiment, a significant increase in the SCL-90-R scores was observed, which was classified as a depressive tipping point \cite{Wichers_2016}. As a consequence of this increase, the participant and his psychiatrist decided to resume the use of antidepressants several weeks after the experiment ended.

\subsection*{Item selection and preprocessing}
Not all 76 EMA items were considered for the purpose of order parameter identification. In this context, the momentary items were most suitable due to their high temporal resolution. For reasons of uniformity, the daily items were not incorporated into the analysis. The weekly SCL-90-R items were only used in the form of the averaged item score in order to define the region in which the tipping point occurred, and were not considered individually. Furthermore, momentary item categories that involved numerical categorisation or had conditional requirements (i.e. social company, activities and events) were omitted. This resulted in a selection of 28 items to be considered for the order parameter analysis.

\begin{figure}[!ht]
\includegraphics[width = 0.9\linewidth]{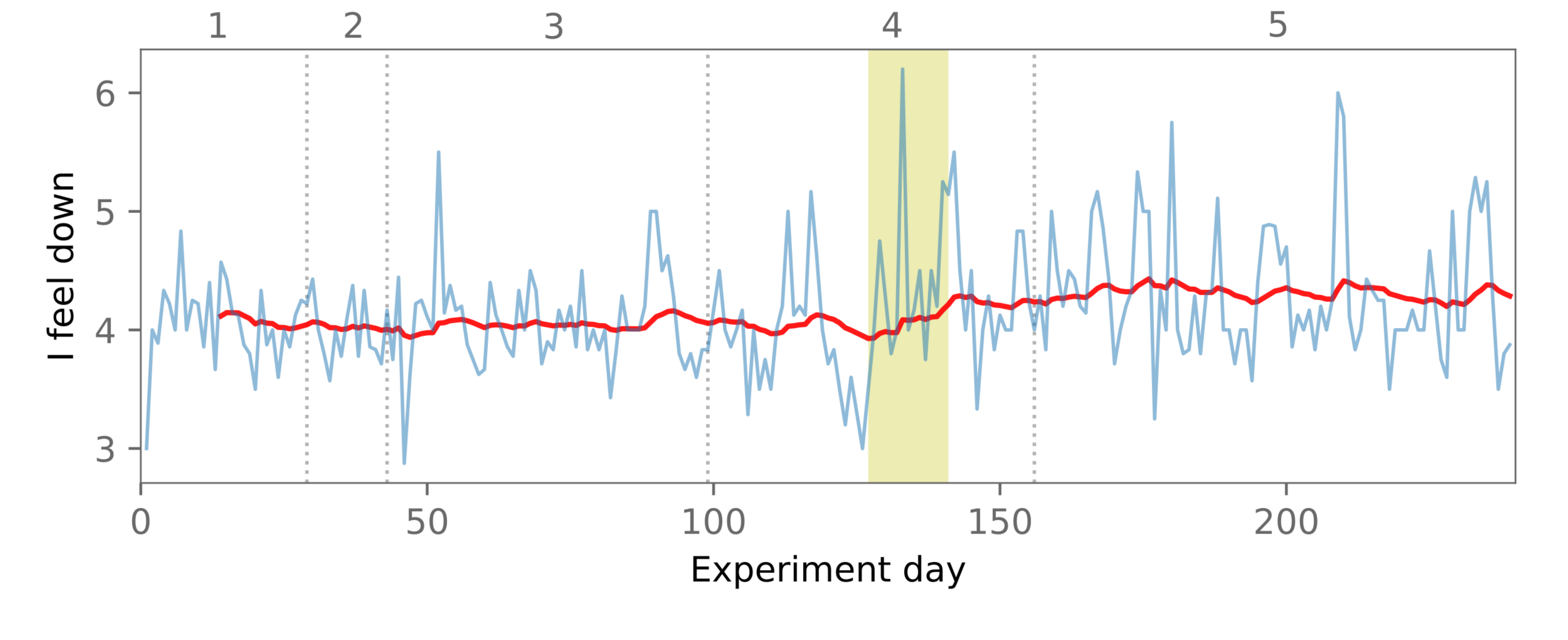}
\caption{{\bf Item preprocessing.}
Example of an exponentially weighted moving average with a half life of 14 days (red) taken from the momentary item `I feel down' after averaging the item scores per day (blue). The dashed lines demarcate the five experimental phases as outlined in the text; the yellow shaded region highlights the region in which the tipping point was identified by \cite{Wichers_2016}.}
\label{fig:item_ewma}
\end{figure}

First, the selected momentary items were averaged for each measurement day. This was done to make sure all measurement points are equidistant for the rest of the analysis. Furthermore, in order to make the data more legible and to assess long-range trends, an exponentially weighted moving average (EWMA) was taken. The half life of the exponential, determining the time scale of the fluctuations that are filtered out by this operation, was taken to be 14 days as a first estimate. This estimate was chosen as a relevant time scale because the DSM-5 requires symptom persistence of 14 days in order for the classification of depressive disorder to apply \cite{DSM5}; the consequences of this decision will be further discussed later. An example of an item and its corresponding EWMA is shown in Fig \ref{fig:item_ewma}.

\subsection*{Order parameter criteria and operationalization}
The aim of the analysis is to propose a general approach towards identifying order parameter related properties of a system undergoing a tentative transition, by systematically applying relevant criteria. Note that we do not claim to determine the exact order parameter itself, which we expect to be a more fundamental quantity that is likely inaccessible through these data alone. Rather, we attempt to identify elements of the dataset that are most likely to be coupled to a hypothetical order parameter.

To assess potential correlation between each item and the order parameter, two basic criteria were verbally formulated as follows:

\begin{enumerate}
\item The item undergoes a noticeable \emph{shift} over the course of the tentative transition.
\item The item displays \emph{increased fluctuations} prior to the tentative transition.
\end{enumerate}

\noindent These criteria were chosen to reflect the most fundamental properties of Landau theory. Symmetry breaking, which could be considered a third key feature of a Landau phase transition, was not considered due to difficulties in quantification of this property. Broken symmetry can be especially difficult to identify at non-equilibrium transitions, where it can manifest in a variety of ways \cite{Haken_1975}. The two stated criteria were quantitatively operationalized to allow for systematic testing of the available data.

\subsubsection*{1. Shift}
The presence of a shift was assessed through comparison of the mean EWMA level after the tipping point (day no. $\geq$ 141) to the mean EWMA level in the baseline period prior to antidepressant reduction (phases 1 \& 2, day no. $<$ 43). If a shift was registered that exceeded the standard deviations of both levels, the criterion was met. Note that this operationalisation also includes items that display a gradual shift occurring over the full course of the medication reduction rather than a distinct shift at the tentative transition point. Although in an ideal case a sharp and clean increase may be expected, an order parameter (or quantities related to that order parameter) can also show more smeared behaviour depending on the nature of the transition, the level of homogeneity in the system or the tuning parameter that is driving the transition.

\begin{figure}[!h]
\includegraphics[width = 0.9\linewidth]{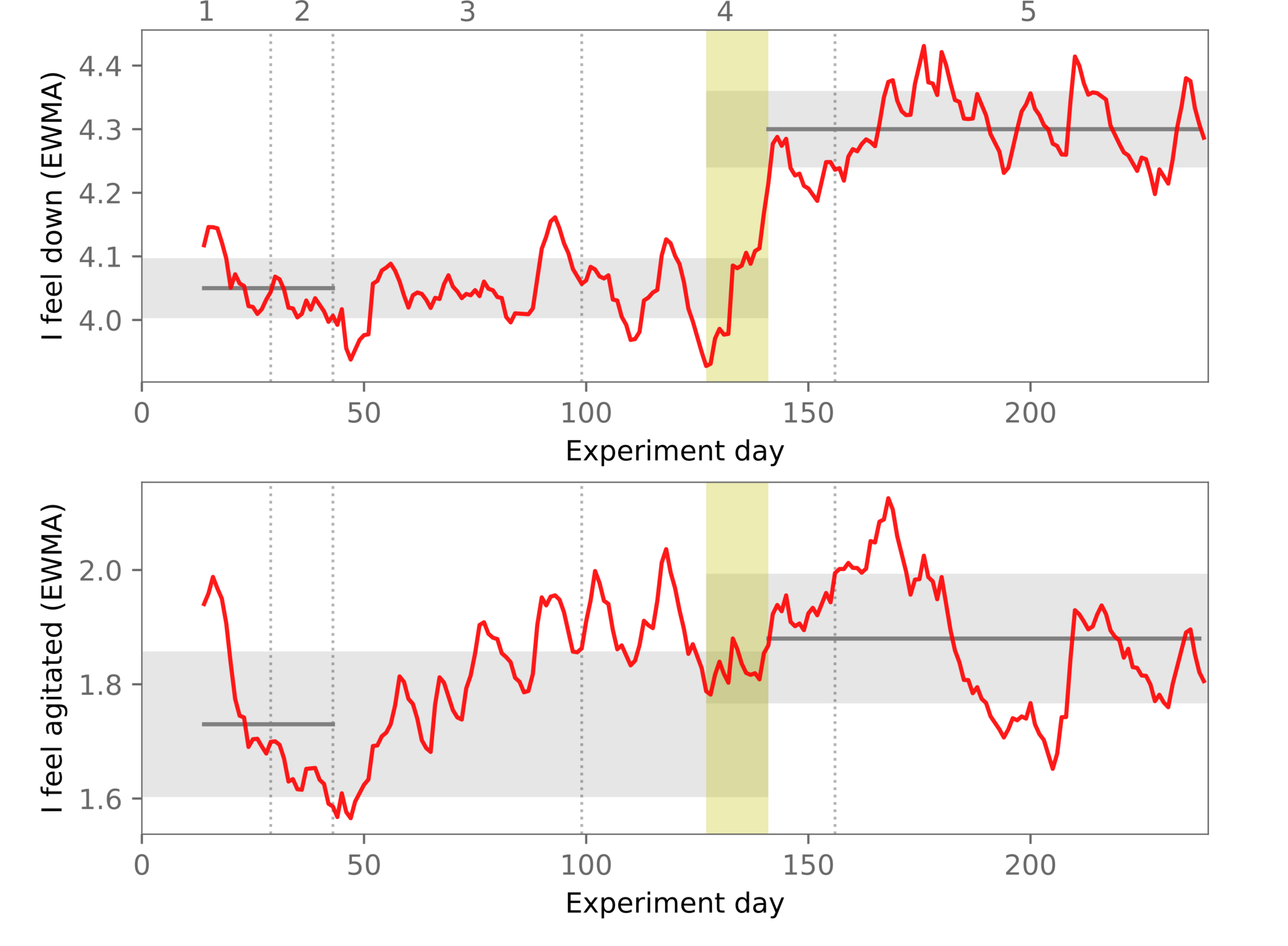}
\caption{{\bf Criterion 1 - Shift.}
Representative examples of the assessment of criterion 1 (the presence of a shift) in two momentary items. The horizontal grey lines indicate the baseline and post-transition means. The grey shaded areas indicate the corresponding standard deviations. Looking at the standard deviations, the criterion was met for the item in (\textbf{a}), but not for the item in (\textbf{b}).}
\label{fig:item_ewmashift}
\end{figure}

The procedure is illustrated in Fig \ref{fig:item_ewmashift}. Both upward and downward shifts were considered; positive (negative) items displaying an upward (downward) shift were manually excluded.

\subsubsection*{2. Fluctuations}
In a (continuous) phase transition, the order parameter is expected to show increasingly large fluctuations that extend through the entirety of the system at the moment of the transition. Therefore, we expect to see heavier fluctuations before the transition than in the baseline period or after the transition in items that are related to the order parameter.

The fluctuations were first detrended to remove the effect of long-term behaviour (such as the shift from the first criterion). This was achieved by subtracting the EWMA from the item scores averaged per day. This means that effectively only fluctuations on a timescale below 14 days were considered.

\begin{figure}[!ht]
\includegraphics[width = 0.9\linewidth]{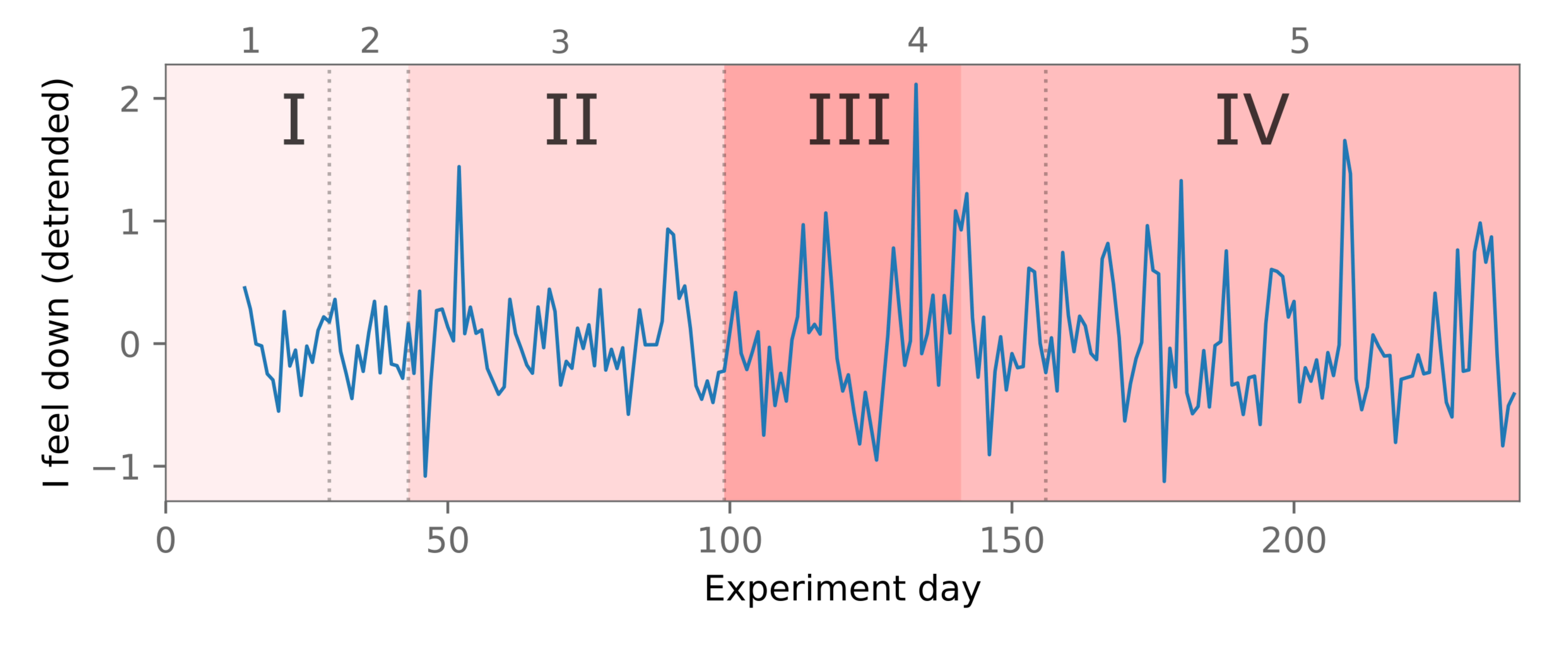}
\caption{{\bf Criterion 2 - Fluctuations.}
Example of the assessment of criterion 2 (fluctuation increase) in the detrended fluctuations (item score per day minus EWMA) of a particular momentary item. The red shaded areas indicate the different regions I-IV as described in the text; the intensity of the colour corresponds to the magnitude of the variance in that region. Since the evolution of the variance fulfills the requirement from equation \ref{eq:fluct_crit}, the criterion was met for this item.}
\label{fig:item_fluct}
\end{figure}

The items were then divided into four regions: (I) the baseline region (phase 1 \& 2, day no. $<$ 43, as above), (II) the region of antidepressant reduction (phase 3, $ 43 \leq$ day no. $<99$), (III) the region prior to and during the transition where the concentration was zero (part of phase 4, $ 99 \leq$ day no. $<141$) and (IV) the region after the transition (rest of phase 4 and phase 5, day no. $\geq141$). The criterion of fluctuation increase was then met if:

\begin{equation}
\label{eq:fluct_crit}
\mathrm{Var(I)} < \mathrm{Var(II)} < \mathrm{Var(III)} > \mathrm{Var(IV)}
\end{equation}

\noindent thus indicating an increase in variance prior to and during the tentative transition. Note that the variance in this case does not correspond to noise but instead captures the true fluctuations of the relevant quantity, since the scores directly reflect changes in the actual mental or physical state of the participant. The procedure is illustrated in Fig \ref{fig:item_fluct}.

\subsection*{Qualitative methods}
Apart from the practical feasibility of order parameter identification techniques in systems outside of the context of physics, this study also aims to address the potential added value of this interdisciplinary translation of methods to both scientific and clinical practice. As a means to contextualise the quantitative results from the data analysis, the quantitative methods were supplemented by an exploratory qualitative study in the form of a small number of interviews. This qualitative component should be interpreted as a first exploration into the topic, highlighting questions and areas of interest that serve to provide context to the analysis presented here and can be more thoroughly pursued in later work.

The aim of the interviews was to make preliminary advancement towards (1) characterizing the position of the order parameter analysis within the broader context of complexity science, (2) identifying relevant motivations and barriers for researchers to apply interdisciplinary methodology in their work, and (3) exploring potential avenues for practical application of phase transition modelling. For this purpose, a number of researchers active in the field of complexity science were interviewed. The decision to perform individual interviews for this purpose was made based on the need for rich, person-specific information with a high level of detail \cite{Jensen_2016}.

A total of four participants were selected for the interviews by purposive sampling \cite{Marshall_1996} based on their affiliation with complexity science, as apparent from public statements and history of publications. Participants were selected with a background in either medical science ($n=2$) or behavioural science ($n=2$) to allow for sufficient intercomparability (see S1 Table. Participant information). Potential participants were chosen through the network of this study's collaborators and approached via email by the first author (MMG). All invited participants agreed to participate in the study. Participants had no detailed knowledge about the purpose of the study and were only given a high-level introduction before the interview to avoid bias and allow for open responses. The interviews were all conducted one-on-one over video call using Zoom, allowing the participants to join from either a professional or private environment.

The interviews were conducted by the first author (MMG), a MSc student in Science in Society with training in qualitative methods, with whom the participants had not previously been acquainted. She performed the interviews according to a flexible interview guide (semi-structured) that was composed based on the three objectives described above and contained sections on the participants' interpretation of complexity science, their experience with interdisciplinary methodology and collaboration and the practical relevance and applicability of their work (see S1 Text. Interview guide). The interview guide was provided with feedback by the co-authors and tested in a trial interview with a collaborator. The duration of each interview was between 50 and 60 minutes. All interviews were conducted in Dutch, the native language of the researcher performing the interviews and three out of four participants. The non-native speaker was given the choice between Dutch and English and opted for the former. Both audio and video recordings were made of each interview after the participants provided their verbal consent; no additional notes were taken during the interviews.

The audio recordings were transcribed nearly verbatim (excluding immediate repetitions and common filler words, but maintaining interjections and grammatical errors) by the researcher who performed the interviews (MMG). Participants were presented with a summarised interpretation of their transcript shortly after the interviews, to which three out of four participants provided feedback and corrections. Transcripts were analysed using ATLAS.ti and organised into emerging themes using a codebook that was established by a combination of inductive and deductive methods (see S2 Table. Codebook). No second coder was consulted at this stage of the study. Given the limited number of interviews performed within the scope of this work, data saturation could not be achieved. However, a relatively high degree of thematic overlap emerged from the participants' responses (for code frequencies, see S3 Table. Codebook). Individual responses were anonymised for the purpose of this report.

Insights from the interviews will be discussed after the quantitative results.

\section*{Results}
\subsection*{Order parameter analysis}
All 28 selected items were subjected to the two criteria as formulated in the Methods section. The selected items and the results from each criterion are shown in Table \ref{tab:criteria}. Full plots of all items, including their EWMA, are available in the Supporting Information (see S1 Fig. All considered items). Starred items did show a shift that met criterion 1, but the shift was downward (upward) for positive (negative) items and thus manually excluded. The items to which this applies appear to be related to the participant feeling increasingly energetic, which might be a consequence of declining side effects of the medication \cite{Fava_2006} rather than being inherent to the depression itself. Items adhering to both criteria are printed in bold.

\begin{table}[ht!]
\centering
\begin{tabular}{lcc}
\toprule
Item name                & \multicolumn{1}{l}{1. Shift} & \multicolumn{1}{l}{2. Fluctuations} \\ \midrule
I feel relaxed           & X                                     &                                            \\
\rowcolor[HTML]{EFEFEF}
\textbf{I feel down}     & X                                     & X                                          \\
I feel irritated         & X                                     &                                            \\
\rowcolor[HTML]{EFEFEF}
I feel satisfied         &                                       &                                            \\
\textbf{I feel lonely}   & X                                     & X                                          \\
\rowcolor[HTML]{EFEFEF}
\textbf{I feel anxious}  & X                                     & X                                          \\
I feel enthusiastic      &                                       &                                            \\
\rowcolor[HTML]{EFEFEF}
I feel suspicious        & X                                     &                                            \\
I feel cheerful          &                                       & X                                          \\
\rowcolor[HTML]{EFEFEF}
\textbf{I feel guilty}   & X                                     & X                                          \\
I feel indecisive        &                                       & X                                          \\
\rowcolor[HTML]{EFEFEF}
I feel strong            &                                       & X                                          \\
I feel restless          & X                                     &                                            \\
\rowcolor[HTML]{EFEFEF}
\textbf{I feel agitated} & X                                     & X                                          \\
I worry                  & X                                     &                                            \\
\rowcolor[HTML]{EFEFEF}
I can concentrate well   & *                                     & X                                          \\
I like myself            &                                       &                                            \\
\rowcolor[HTML]{EFEFEF}
I am ashamed of myself   & X                                     &                                            \\
I doubt myself           & X                                     &                                            \\
\rowcolor[HTML]{EFEFEF}
I can handle everything  &                                       & X                                          \\
I am hungry              & X                                     &                                            \\
\rowcolor[HTML]{EFEFEF}
I am tired               & *                                     & X                                          \\
I am in pain             &                                       &                                            \\
\rowcolor[HTML]{EFEFEF}
I feel dizzy             &                                       &                                            \\
I have a dry mouth       &                                       &                                            \\
\rowcolor[HTML]{EFEFEF}
I feel nauseous          & X                                     &                                            \\
I have a headache        &                                       &                                            \\
\rowcolor[HTML]{EFEFEF}
I am sleepy              & *                                      &                                            \\ \bottomrule
\\
\end{tabular}

\caption{\textbf{Order parameter analysis.} Results from testing the selected items against the formulated order parameter criteria. If a criterion was met, the item is marked with an X. Starred items (*) did show a shift, but the shift was downward (upward) for positive (negative) items and thus excluded from meeting the criterion. Bold items meet both criteria.}
\label{tab:criteria}
\end{table}

The results of the order parameter analysis may depend on the timescale of the EWMA smoothing applied to the data. This timescale is determined by the averaging half life of the EWMA, which so far has been set to 14 days. Therefore, the robustness of the results with respect to this parameter should be assessed.

To this end, all selected items were again subjected to the two order parameter criteria, but this time the EWMA half life was varied between 5 and 80 days. This means that, for longer half lives, a) the EWMA used to assess the presence of a shift undergoes more smoothing, and b) fluctuations of longer timescales are taken into account, since the detrending procedure determines the upper limit to this timescale. In order to assess the outcomes of the two criteria as a function of EWMA half life, they were quantified as follows. The relative shift $\Delta$EWMA compared to the baseline was defined as

\begin{equation}
\Delta \mathrm{EWMA} \coloneqq \frac{\mu(t \geq 141) - \mu(t < 43)}{\mu(t < 43)} \cdot 100\%
\end{equation}

\noindent with $\mu$ the mean of the EWMA in the indicated region and $t$ the time in days.

The relative variance increase $\Delta$Var compared to the average of the baseline variance and the variance after the transition was defined as

\begin{equation}
\Delta \mathrm{Var} \coloneqq \frac{\mathrm{Var(III)} - \frac{1}{2}(\mathrm{Var(I) + Var(IV)})}{\frac{1}{2}(\mathrm{Var(I) + Var(IV)})} \cdot 100\%
\end{equation}

\noindent with Var(X) the variance in region X, as established in the definition of the fluctuation criterion.

\begin{figure}[!ht]
\includegraphics[width = 0.9\linewidth]{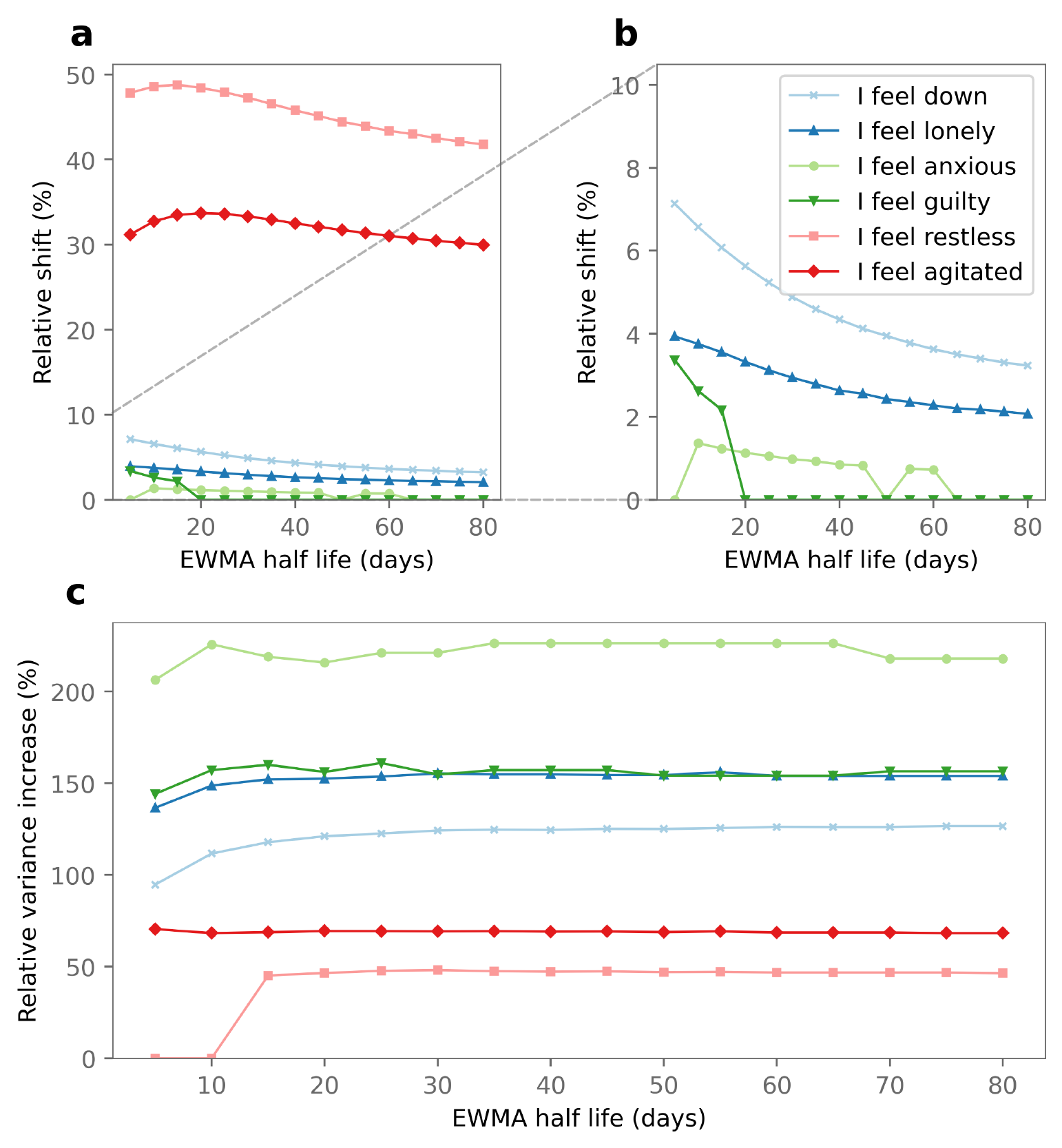}
\caption{{\bf Timescale dependence.}
Half life dependence of the relative shift (\textbf{a} + \textbf{b}) and relative variance increase (\textbf{c}) for items meeting both criteria for at least one half life value. Fig (\textbf{b}) shows a zoomed in view of the items with a smaller relative shift. Values are set to zero if their respective criterion is not met. Varying the half life primarily affects which items meet criterion 1 (related to the relative shift), whereas criterion 2 (related to relative variance increase) remains largely unaffected.}
\label{fig:hl_sensitivity}
\end{figure}

Both $\Delta$EWMA and $\Delta$Var were set to zero if their respective binary criterion (see Methods section) was not met. Their definitions are not meant to have quantitative significance by themselves, but rather serve to provide a comparative measure to help visualise the half life parameter dependence. $\Delta$EWMA and $\Delta$Var were then calculated for each of the included items, for half lives ranging between 5 and 80 days. Items meeting both criteria for at least one half life value were plotted. The results are shown in Fig \ref{fig:hl_sensitivity}.

Fig \ref{fig:hl_sensitivity} illustrates how some items can go from not meeting one of the criteria to meeting it (i.e. changing from zero to non-zero) depending on the chosen value for the EWMA half life. Nevertheless, the procedure shows that the dependence of the criteria on the EWMA half life is only weak for most items. The relative shift corresponding to criterion 1 shows a slight decrease with increasing half life, which is to be expected due to the enhanced `smearing' of the transition with heavier averaging. The relative variance increase corresponding to criterion 2 seems largely unaffected by a change in half life, meaning that considering fluctuations at larger time scales does not change the relative height of the peak nor does it have much effect on which items are meeting the threshold. The items `I feel down', `I feel lonely' and `I feel agitated' were most stable with respect to the parameter change; they met both criteria for the entire considered range. The shift in the item `I feel guilty' was no longer considered large enough to meet criterion 1 for half lives longer than 15 days, and the shift in the item `I feel anxious' only meets the criterion for certain half live values up to 69 days. Finally, the item `I feel restless', which was not previously identified as an order parameter candidate, was found to consistently meet both criteria for half lives of 15 days and up.

The choice of half life therefore mildly affects the items that are identified as order parameter candidates by the two criteria. In practice, the relevant timescale should be determined by knowledge about the underlying process that generated the data; this discussion will not be developed here. Nevertheless, it should be noted that the criteria in their current form do not allow for intermediate outcomes (e.g. partially meeting a criterion) and can thus lead to unintended exclusion of possibly relevant items. To avoid this, it seems sensible to consider a range of timescales rather than a single fixed timescale.

\subsection*{Insights from the interviews}
The contextual insights obtained from the exploratory interviews will be presented generally and interpreted in relation to the order parameter analysis proposed here. The discussion will be structured along three main axes:

\begin{enumerate}
\item Understanding the position of the phase transition approach in the diverse landscape of complexity science.
\item Identifying motivations and barriers for using interdisciplinary methods such as phase transition modelling in complexity science.
\item Exploring the possible practical relevance and distance to clinical application of interdisciplinary methods such as phase transition modelling in complexity science.
\end{enumerate}

\noindent Supporting quotes from the transcripts are translated to English by the first author; the original Dutch wording can be found in the Supporting Information (see S2 Text. Original interview quotes). The quotes are edited only for legibility and do not leave out substantive information. All edits are made explicit by square brackets. Participants are labelled 1--4, with participants 1 and 3 coming from a behavioural science background and participant 2 and 4 from a medical background (see S1 Table. Participant information).

\subsubsection*{1. The complexity science landscape}
Although all participants considered their research to be related to complexity science, they maintained different interpretations of what complexity means. Several characteristics came up repeatedly, but were given different weights depending on the participant. Multicausality, strong interactivity and dynamic/temporal evolution were most commonly mentioned as core features of complex systems, closely followed by non-linearity of input-output relations. Some participants also mentioned the need to consider complex systems at the level of individuals rather than group averages, since individual dynamics can get lost when aggregated. Participant 3 phrased it as follows:

\begin{displayquote}
``...so I think the most important difference is to [...] analyse first and then aggregate, so in principle we start from [a] case, and [...] if we have multiple cases, then we look for regularities instead of throwing everything on one pile and looking for connections then."
\end{displayquote}

Moreover, there was an apparent dichotomy in the way the participants operationalized complexity. For some, complexity was associated with quantitative, data-driven methods. Other participants related complexity to a qualitative, systemic approach (also referred to as ``systems thinking"). Participants on the quantitatively oriented end were most explicit about the specific methods or tools they did and did not associate with complexity science, whereas for the qualitatively oriented participants, more emphasis was laid on complexity as a conceptual framework.

In general, all participants contrasted the complexity-related approach to traditional research methods that they consider most common in their field. Although some mentioned that they can be used alongside each other, methods originating from complexity are positioned as inherent counterparts to regular `evidence-based practice', i.e. the gold standard of clinical decision-making based on typical linear, statistical methods. Some participants therefore referred to adopting the perspective of complexity science and its consequences as a paradigm shift.

The order parameter analysis from this work occupies a slightly different position on the spectrum of complexity interpretations. The most frequently mentioned elements of complexity do not necessarily apply: modelling a tipping point as a phase transition does not imply the existence a multitude of causal factors for instance, but instead assumes the existence of an unknown control parameter and subsequently abstracts the individual constituents of the system in an attempt to identify macroscopic order. Moreover, the study of complex systems in physics is not a distinctive or alternative approach compared to `traditional' physics research, in contrast to the participants' view on complexity science. Therefore, the interpretation of phase transitions in complex systems in physics might be different to how scientists view complex systems studies in other areas of science.

\subsubsection*{2. Motivations and barriers for interdisciplinarity in complexity science}
By far the most frequently mentioned motivation for adopting the complexity science perspective in general are the perceived shortcomings of traditional `non-complex' approaches. Participant 4 for instance described a research line that had failed to produce substantial results in several decades due to an overly narrow view on causality. They then comment:

\begin{displayquote}
``...so then you need a very different kind of paradigm and you also need different methods to escape from yet another simple linear statement that runs the risk of basing another forty years of research on one mechanism."
\end{displayquote}

All participants recalled examples of research lines that either left out important elements in favour of a linearised view or focused on standardised approaches that lacked intrapersonal effectiveness, which led these studies into useless conclusions or dead ends. The participants indicated that the complexity perspective better fitted the reality or practice of their object of study, allowing for new realisations or insights that could not have been achieved within the traditional paradigm.

The fact that complexity science appears to occupy such a distinctive niche in the scientific world comes with a few challenges. Participants mentioned particular difficulties with the lack of acceptance of the complexity perspective by other areas of science. Because complexity provides an alternative conceptual view or even a replacement of traditional methods, recognition of complexity methods within more mainstream science was said to require a substantial paradigm shift.

Interdisciplinary methods and collaborations appear to be commonplace within the field of complexity science. Participant 2 in particular, and to a lesser extent also participant 1 and 4, mentioned that an interdisciplinary approach is inherent to their type of research and their operationalisation of complexity. For participant 1, this implied making use of expertise that is not naturally available within psychological science, such as computational modelling. Participant 2 and 4 on the other hand described interdisciplinarity as the incorporation of a range of different stakeholder groups and medical disciplines into their research, which relates to the large weight of multicausality in their respective interpretations of complexity. Participant 2 phrased it as follows:

\begin{displayquote}
``For me that is also a characteristic, [...] the types of questions that you get from such a complexity perspective are just very interdisciplinary."
\end{displayquote}

The most important motivations for pursuing these interdisciplinary efforts came down to the added value of additional expertise and the possibility to obtain new insights, similar to the motivation for engaging with complexity science in general. This ties in to the previous observations on how the participants pointed at the failures and limitations of traditional techniques from within their own discipline, after which they chose to look out for methods across the disciplinary borders. Participants therefore expressed the need for dialogue with experts from other scientific disciplines to collectively interpret problems and questions from a different angle and improve mutual understanding. For example, participant 3 describes the added value of interdisciplinary collaboration as follows:

\begin{displayquote}
``...I'm also just a psychologist, and I have read some literature, but I don't feel skilled enough to properly understand all of these things. On the other hand, physicists don't have a clue what we are talking about. So I [...] think especially this interdisciplinary collaboration and testing with each other [...] what we mean and [...] what we would expect, [...] I would find that immensely valuable."
\end{displayquote}

The order parameter analysis proposed here might present such an alternative angle based on established theory in physics. Phase transition modelling provides a way to conceptualise and understand complex behaviour outside of physics, in this case a depressive tipping point, in a different way. In that regard, phase transition modelling could be an alternative avenue to pursue for cases where new interdisciplinary methodologies are required.

Interdisciplinary collaborations are also subject to potential barriers. Firstly, all participants had experienced communicative issues in some form during their interdisciplinary research. The participants all had their own associations and implications related to complexity and without explicit exposition of these differences, misunderstandings can easily arise. Differences occurred on many levels, from the interpretation of concepts to the comparison of results obtained by different methods. For example, participant 4 recalled a collaboration with management scientists who implicitly referred to the concept of `resilience' on the level of organizations rather than individuals, which did not surface until much further into the project. Participants stressed the necessity to make dissimilarities like these as explicit as possible in advance, to avoid unwanted surprises midway through a collaborative effort.

Moreover, some participants commented on their difficulties translating the methodologies and concepts they use from other disciplines to their own. Sometimes metaphorical similarities between fields are easy to identify, but it can be more challenging to translate these to concrete research methods which can then be tested and empirically verified. Due to the fact that complexity science is a relatively young discipline, participants indicated that the effectiveness or adequacy of complexity research compared to traditional methods is not yet established. Many studies are still in a relatively conceptual phase and have only recently become more analytically applied, which means that their added value to e.g. clinical interventions is not certain. Participant 1 in particular expresses caution in this regard, emphasising the importance of being willing to abandon a new research direction if it cannot be proven to be more effective than existing approaches. Participant 3 comments:

\begin{displayquote}
``...we really still have to see to what extent all of these metaphors [...] are really suitable for us, and whether it all really works the same."
\end{displayquote}

Phase transition modelling is therefore likely subject to differences in assumptions underlying complex systems studies in physics and behavioural science. The differences that emerged in this work and their consequences will be addressed in more detail in thedetail the discussion.

\subsubsection*{3. Practical relevance and clinical application of interdisciplinarity in complexity science}
The relative focus on applicability versus fundamental understanding differed for each participant. Participant 3, whose research line is the most relevant to the phase transition approach presented here, has noticed an intuitive match of their conceptual ideas with therapists in clinical practice. They also mentioned that the phase transition metaphor in particular resonates well with clinicians, who recognise the idea of periods of instability or fluctuations that precede significant clinical change. However, they urged caution in this respect, stressing the importance of empirical and quantitative backup to these conceptual claims.

Another point that was brought up by both participant 3 and 4 was the increasing role of uncertainty that comes with the practical implementation of complexity methods. This relates to the limited predictive horizon that is often associated with complex systems, in contrast to a more traditional deterministic, standardised view. According to participant 3, the uncertainty of intervention effects calls for continuous monitoring and the ability to adjust based on individual responses. Participant 4 argued that definite answers and calculable outcomes can simply not be achieved in the case of complex systems, and therefore the inherent uncertainty should be embraced and incorporated into decision making rather than hidden in heavily averaged numbers. From this perspective, phase transition theory might be unlikely to produce traditionally applicable results in terms of interventions or treatment methods with a high degree of outcome certainty. Rather, it could provide insights into the underlying mechanism of a process on an intra-individual level.

\section*{Discussion}
The order parameter focused phase transition modelling approach presented here is different from other existing methods of studying tipping points on several fronts: it does not attempt to identify a tipping point or phase transition \cite{Schiepek_2020,Viol_2022}, rather to characterise nature of an already recognised tipping point through identification of order parameter related properties; it also does not aim to provide predictive value, in contrast to early warning signals \cite{Scheffer_2009,Olthof_2019}, or other dynamical analyses \cite{Schiepek_2010,Schiepek_2016}, but instead offers an alternative conceptualisation and way of viewing psychopathology. Moreover, although fluctuations and variance have commonly been associated with tipping points \cite{Litt_2001,Smit_2019}, the explicit connection to other characteristics of order parameter behaviour (i.e. the shift criterion) has not been made before to the best of our knowledge.

Several points of discussion should be addressed. First of all, as emphasized earlier, the order parameter analysis proposed here does not claim to result in an exact identification of the order parameter of this particular transition. Instead, it provides suggestions about the types of symptoms that might be most closely related to an order parameter, under the hypothesis that such an order parameter exists; these symptoms might be interpreted as particularly relevant to the depressive transition in this individual. It should be noted that another core feature of phase transitions has thus far not been addressed: symmetry breaking, a reduction in the symmetry of a system (translational, temporal, gauge etc.) over the course of a phase transition due to the macroscopic ordering in the post-transition phase \cite{Hohenberg_2015}. Additional investigation from the perspective of behavioural science and theorizing of psychopathology are required for further interpretation of the results.

Secondly, potential extension of the results from this $n=1$ study to other cases of clinical tipping points requires some important considerations. As also suggested by the supplementary interviews, the complexity science perspective in psychology puts emphasis on individualisation and personalisation of psychological conditions and their corresponding symptoms, due to the fact that psychological systems appear to violate certain properties required for a valid statistical analysis on aggregated groups \cite{Olthof_2020}. From this paradigm, one might expect that the specifics of the symptomatic structure and corresponding order parameter of a depressive transition are different to each individual. At the same time, classical theory of phase transitions and its subsequent translation into other complex systems relies on universality, suggesting that microscopic differences should not play a role in a macroscopic description. One solution might be to adopt a higher-level interpretation of universality: rather than expecting a universal set of symptoms related to an order parameter that is the same across all individuals, we only assume the existence of certain universal features of phase transitions that occur in physics and behavioural science alike. This validates the idea that core symptoms can be identified over the course of sudden clinical change by means of order parameter analysis, while simultaneously not implying generalised or population-wide uniformity. Nevertheless, repetition of the analysis on other datasets would be highly valuable.

\begin{figure}[!ht]
\includegraphics[width = 0.9\linewidth]{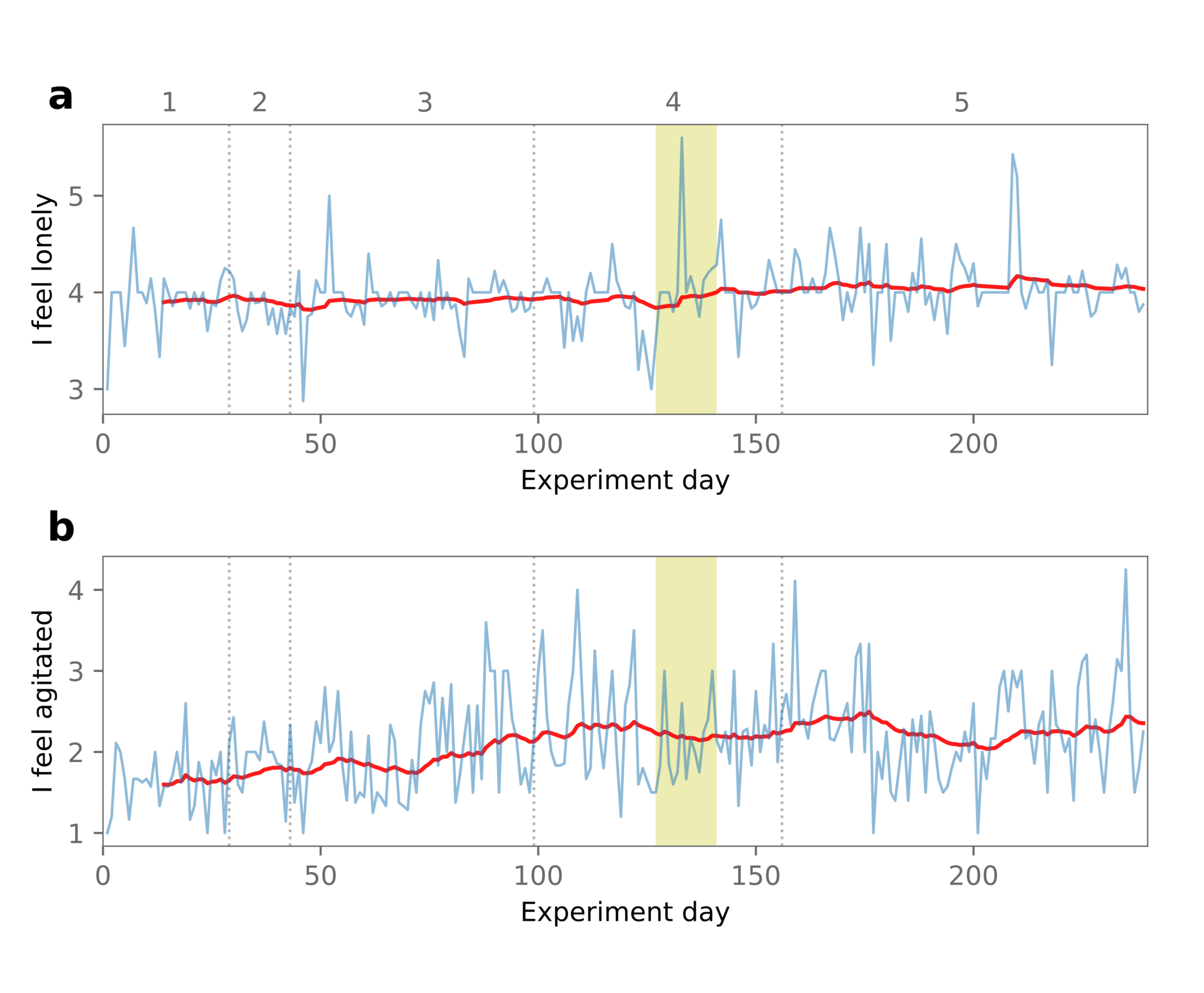}
\caption{{\bf Example of qualitatively different behaviour between items identified as order parameter related.}
Comparison of items (\textbf{a}) feeling lonely and (\textbf{b}) feeling agitated. Both items consistently meet both criteria. Notice how the relatively small increase in item scores occurs within the yellow shaded region in (\textbf{a}), while the increase is much more gradual and larger in magnitude in (\textbf{b}).}
\label{fig:item_diffs}
\end{figure}

Moreover, Fig \ref{fig:hl_sensitivity} seems to suggest possible qualitative differences between the items found to meet the order parameter criteria. The items can be roughly divided into two categories: items with a larger shift and a smaller variance increase (`Restless', `Agitated') and items with a smaller shift and larger variance increase (`Down', `Lonely', `Anxious', `Guilty'). Upon inspection, there appears to be a difference in the evolution of these items over time. Fig \ref{fig:item_diffs} illustrates this difference by comparing the item `I feel agitated' to the item `I feel lonely'.

The items from the first category, such as `I feel lonely', have a shift that is relatively small, but also sharp. The items from the second category on the other hand, such as `I feel agitated', show a much larger shift that occurs more gradually over time. Although both groups of items thus meet the formulated criteria, they may be driven by different mechanisms or coupled to the order parameter in different ways.

Another possibility is that the items in the second category are not related to an order parameter, but rather to an unknown control parameter driving the tentative depressive transition. The presence of a shift can only be assessed as a function of time by implicitly assuming the existence of some unidentified control parameter that more or less monotonically increases over time. The exact nature of the control parameter is unknown, although it is assumed to be related to the antidepressant medication dosage \cite{Wichers_2016}. Feeling `agitated' and `restless' might thus not be related to an expression of the order in the system, but instead echo the medication reduction as the hypothetical control parameter.

The ignorance about the control parameter of the transition could have more implications. For instance, it has been shown that upon varying a combination of control parameters together, the behaviour of a system around a critical point may change \cite{Dai_2015}. In particular, the increase in fluctuations typically associated with critical slowing down might not be observed despite an imminent transition \cite{Titus_2020}. Since the exact drivers of the transition in our case study are unknown, it is possible that this effect masks or alters the fluctuation characteristics of some items around the tipping point.

In terms of added value, the supplementary interviews suggest that researchers in some areas of science are looking for new ways to address certain problems in their field and seek inspiration and expertise beyond the disciplinary boundaries, in which respect phase transition modelling could prove a promising approach. However, it should be carefully considered that the way in which physics interprets properties of complex systems can distinctly contrast with the conceptions that exist within other areas of complexity science, which could lead to difficulties with communicating and aligning views. Predictable or universal behaviour is common in exact sciences like physics, but that paradigm does not automatically translate to more human-oriented fields like psychology. The apparent discrepancy between universality in macroscopic descriptions and the pursuit of individuality before aggregation, as discussed before, is one such example.

The practical applicability of phase transition modelling in a clinical setting is difficult to establish at this stage. Although the conceptualisation might metaphorically match well with clinical intuition, translating this insight into practical added value is yet to be achieved. Based on the preliminary results presented here, analysing a clinical transition in terms of order and disorder could help highlight symptoms that are particularly strongly coupled to a certain type of transition, which could provide potentially valuable treatment targets in the future (similar to what is attempted by certain studies based on network analysis \cite{Borsboom_2013,Contreras_2019}) and help limit the number of elements required to characterise disease progression. It might also shed light on similarities and differences between the nature of the potential transitions occurring in people with other forms of psychopathology.

Other potential applications that might be proposed relate to predicting clinical tipping points. An increase or change in the structure of fluctuations, similar to what has been studied here, has often been associated with predicting critical transitions \cite{Scheffer_2009}. However, the added value of the phase transition approach in terms of predictive power is only little compared to established methods. Rather, it aims to advance fundamental understanding and conceptualisation of tipping points in complex systems and can provide one of the alternative perspectives on psychopathology that complexity-oriented behavioural scientists are looking for. Despite these caveats, the analysis presented here could aid in identifying target observables for monitoring programs, for instance in preventing relapse. In this way, the order parameter candidates could provide suggestions for relevant symptoms to track in a particular individual.

Finally, from an experimental view, the results of a phase transition based analysis such as presented here could potentially be a useful asset in the design of future momentary assessment studies. Extensive longitudinal experiments for obtaining time series data can be incredibly demanding in terms of the time investment and dedication required from the participants. If further study into this area could highlight certain items as particularly relevant or irrelevant to a certain transition, some of the burden might be alleviated by reducing the number of items necessary to understand the participant's development over time.

\section*{Conclusions}
In this work, we aimed to address the validity of the intuitive similarity between phase transitions in physics and tipping points in complex systems outside of physics, by assessing a case study from psychology concerning a tipping point in depressive symptoms. To this end, an analysis focused on the order parameter was proposed as a first step towards phase transition modelling in systems outside of physics. The formulation of two quantitative criteria reflecting expected order parameter behaviour allowed for the identification of symptoms from the case study that are most likely to be related to the hypothetical order parameter of the depressive transition in this dataset. Supplementing these results, exploratory interviews helped clarify the position and potential added value of this approach within the interdisciplinary landscape of complexity science. The findings suggest that there exists a demand for alternative perspectives to solve complex problems in behavioural and medical science, to which phase transition modelling could provide one possible answer. Additionally however, several subtle interpretative differences between complex systems studies in physics and other scientific areas were recognised, which might limit the translatability of the phase transition approach and should be carefully considered in any future continuation of this attempt.

More study into this area would be valuable. Regarding the results presented here, it would be interesting to further study the relation between the items that came up as order parameter relatives, both by means of quantitative methods as well as qualitative evaluation of possible underlying connections in relation to psychopathology. In this light, repetition of the analysis on comparable datasets would be very valuable. This could also help illuminate the level of individuality of symptomatic structures and therefore contribute to the understanding of the meaning of universality in this context. Moreover, an additional number of interviews into this subject should be performed, to further probe the suggestions raised by the exploratory interviews in this work.

Experimentally, it would be interesting to further investigate the role of possible control parameters that might be driving the depressive transition, since this is a blind spot in the dataset used here. This could take the form of regularly probing relevant biomarkers during the measurement period in order to map out the delayed response to antidepressant reduction and to learn more about the behaviour of a possible control parameter and symptoms coupling to this control parameter. Moreover, having data for the period in which the participant and their doctor decided to reintroduce the antidepressant would also be highly valuable in assessing the inverse behaviour of the transition. Hypothetically, it would also be interesting to vary the rate at which the medication is reduced to see if the transition increases or decreases in sharpness. Such an experiment would have to be performed on a single participant however due to large expected interpersonal differences when it comes to biochemical response, which makes this proposal infeasible in practice. In general, it should be recognised that experiments like these are often difficult or even impossible to perform due to ethical considerations, the large uncertainty in the occurrence of a transition and the substantial effort required from the participant. However, findings and recommendations such as presented here could provide direction for experimental design in the future.

With this study, we took a first step towards modelling tipping points in complex systems as phase transitions. The fact that interpretable results were produced and avenues for additional study emerged implies that this approach contains potential and is worth pursuing further. Moreover, the interviews suggest that there exists a growing need for fundamentally alternative, interdisciplinary views on current problems, to which the phase transition approach could provide one possible answer. New questions emerged related to the nature of interdisciplinarity and the meaning and assumptions behind complexity in physics as opposed to other areas of science, which could shape to what extent and in what way physics-based methods can be applied to social sciences, medical science and the like. The effectiveness and potential for clinical utilisation of highly interdisciplinary translations like phase transition modelling is still largely unknown; nevertheless, further pursuit and investigation of these novel ideas definitely seems promising.

\section*{Acknowledgments}
We would like to thank the members of the Radboud Interfaculty Complexity Hub (RICH), in particular Merlijn Olthof, Ren\'e Melis and Jerrald Rector, for the many fruitful discussions.

\nolinenumbers

%
%
%

\section*{Supporting information}
\textbf{S1 Fig. All considered items.} Plots of all items considered in the order parameter analysis, averaged per day (blue line), including their EWMA (red line).

\noindent \textbf{S1 Table. Participant information.} Details of interview participants and their respective backgrounds.

\noindent \textbf{S1 Text. Interview guide.} Full interview guide used to conduct the interviews, in its original language (Dutch).

\noindent \textbf{S2 Table. Codebook.} Codebook used to analyse the interviews, including a description and the frequency of occurrence of each code.

\noindent \textbf{S2 Text. Original interview quotes.} Interview quotes appearing in the main text in their original language (Dutch).

\noindent \textbf{S3 Text. Full interview transcripts.} Full transcripts of the four interviews in their original language (Dutch).

\end{document}